\documentclass[twocolumn,showkeys,aps,prd,showpacs]{revtex4-1}
\UseRawInputEncoding
\usepackage{graphicx}
\usepackage{pifont}
\usepackage[CJKbookmarks,dvipdfm,colorlinks,linkcolor=red,citecolor=blue]{hyperref}
\usepackage{graphicx}
\usepackage{amsmath, amssymb}
\usepackage{makecell}
\usepackage{multirow}
\usepackage{CJK}
\usepackage{mathrsfs}
\usepackage{bm}
\usepackage{amsmath}
\usepackage{dcolumn}
\usepackage{epstopdf}
\usepackage{dsfont}
\usepackage{amssymb}
\usepackage{tabularx}
\usepackage{array}
\usepackage{float}
\usepackage{color}
\usepackage{epstopdf}
\usepackage{mathrsfs}
\usepackage{extarrows}

\begin{document}

\title{Isovalent alloying assisted  anomalous valley Hall effect in hexagonal antiferromagnetic monolayer}

\author{San-Dong Guo$^{1}$, Liguo Zhang$^{1}$, Xiao-Shu Guo$^{1}$ and Gangqiang Zhu$^{2}$}
\affiliation{$^1$School of Electronic Engineering, Xi'an University of Posts and Telecommunications, Xi'an 710121, China}
\affiliation{$^2$School of Physics and Electronic Information, Shaanxi Normal University, Xi'an 716000, Shaanxi, China}
\begin{abstract}
Exploring combination of antiferromagnetic (AFM) spintronics and anomalous valley Hall  effect (AVHE) is one of the most important questions for valleytronic applications. The key to address this issue is to achieve  spin splitting around the valleys in AFM systems. Here, we propose a possible way for achieving AVHE  in  hexagonal AFM monolayer, which  involves the isovalent alloying.  This can break the combined symmetry ($PT$ symmetry) of spatial
inversion ($P$) and time reversal ($T$),  giving rise to spin splitting.  More specifically, the large spin splitting around the
Fermi energy level owes to $d$ orbital mismatch among these  different
transition metal ions.
Based on first-principles calculations, the proposed way can be verified in out-of-plane AFM   $\mathrm{CrMoC_2S_6}$ monolayer, which possesses spontaneous valley polarization and spitting splitting, providing possibility to realize AVHE. It is also proved that tensile strain  can strengthen the valley splitting and maintain the out-of-plane AFM ordering.
Our works provide an experimentally feasible way for developing AFM valleytronic  devices.
\end{abstract}
\keywords{Valley,  Antiferromagnetism, Isovalent alloying~~~~~~~~~~~~~~~~~~~~~~~~~~~~Email:\textcolor[rgb]{0.00,0.00,1.00}{sandongyuwang@163.com}}

\maketitle

\section{Introduction}
Valleys, as an additional degree of freedom of electrons besides charge and spin, lay the foundation  for applications in information encoding and processing\cite{q1,q2,q3,q4}.  The discovery and successful preparation of rich two-dimensional (2D)
materials promote the development of valley concept due to the large separation in momentum space.
2D transition metal dichalcogenides (TMDs)  possess intrinsic broken inversion
symmetry and large spin-orbital coupling (SOC),  and they are widely recognized as the most representative valleytronic materials  with  a pair of degenerate but inequivalent -K and K valleys in the reciprocal space\cite{q8-1,q8-2,q8-3,q9-1,q9-2,q9-3}.
However, spontaneous valley polarization lacks in these nonmagnetic TMD monolayers, and some external methods can achieve valley polarization, such  as external magnetic field, proximity effect, magnetic doping and light
excitation\cite{v5,q9-3,q9-2,v10}. Although these methods
can  achieve  valley polarization, they have
many shortcomings, such as small valley splitting,  destruction of crystal structure and  limited carrier life.

Compared with external methods, the intrinsic
valley polarization is highly desirable. Fortunately,  the ferrovalley semiconductor   has been proposed\cite{q10}, which possesses
 intrinsic valley polarization and can achieve anomalous valley Hall  effect (AVHE).   The ferrovalley  materials are generally built on ferromagnetic (FM) hexagonal symmetric systems
 with broken spatial inversion symmetry\cite{q10,q11,q12,q13,q13-1,q14,q15,q16,q17,q18}.
Compared to FM materials,  antiferromagnetic
(AFM) materials  possess  the high storage density, robustness against external magnetic field, as well as the ultrafast writing speed\cite{v12}.
 Therefore, realizing  valley polarization and AVHE in AFM materials is more meaningful for valleytronic application.
However, 2D  AFM valleytronic materials
with intrinsic valley polarization and realizable AVHE  are still scarce.
Only a few candidate systems have been proposed, such as   $\mathrm{MnPSe_3}$/$\mathrm{Sc_2CO_2}$ or $\mathrm{Cr_2CH_2}$/$\mathrm{Sc_2CO_2}$ heterojunction\cite{v13,v14} and  Janus $\mathrm{Mn_2P_2X_3Y_3}$ (X, Y=S, Se Te; X$\neq$Y)
monolayers\cite{v15}.

In particular, the presence of spin degeneracy poses a significant obstacle to realize  AVHE in AFM systems.
The out-of-plane external electric field has been used to induce spin splitting in  AFM systems, such as A-type AFM hexagonal monolayer $\mathrm{Cr_2CH_2}$ and tetragonal monolayer $\mathrm{Fe_2BrMgP}$\cite{gsd1,gsd2}. This can be explained by the introduction of layer-dependent electrostatic potential
caused by out-of-plane external electric field,  and the spin order of  spin splitting can be reversed by flipping  the direction of electric field. In fact, the AVHE can be achieved in electric-potential-difference antiferromagnetism (EPD-AFM) without external electric field\cite{gsd1,gsd2}, which can be equivalently replaced by a built-in electric field caused by Janus structure.

\begin{figure}
  \includegraphics[width=8cm]{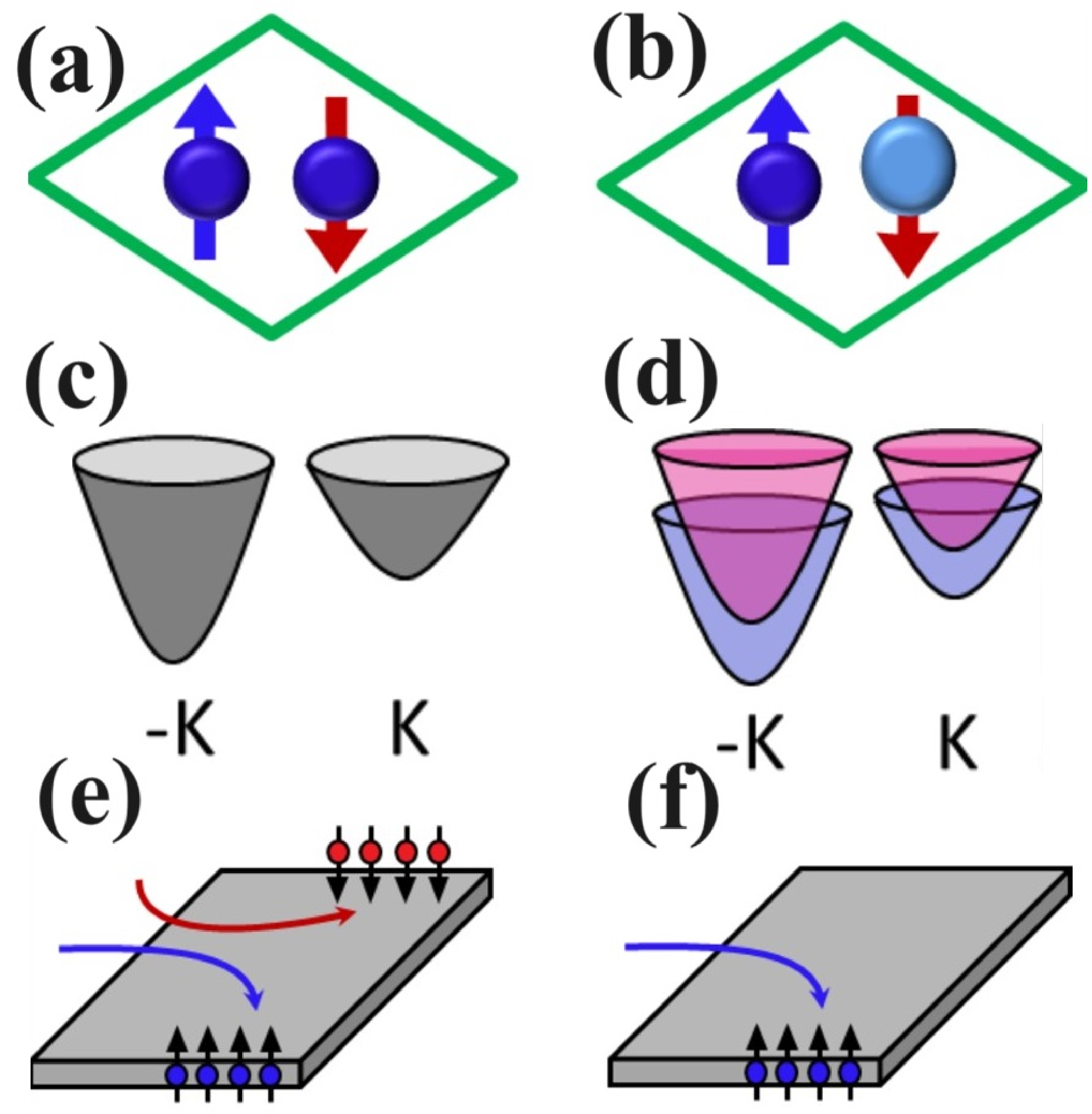}
  \caption{(Color online) (a) and (c): a hexagonal $3d$-AFM monolayer with spin degeneracy  but  nonequivalent  -K and K  valleys; (b) and (d): via isovalent alloying, (a) becomes $3d$- and $4d$-AFM monolayer with removed spin degeneracy, and the  -K and K  valleys still are unequal; In the presence of a longitudinal in-plane electric field,  an appropriate electron doping for two cases  produces  valley-spin hall effect (e) and  AVHE (f). }\label{sy}
\end{figure}

 Here, we propose a way to achieve spin splitting  by isovalent alloying, and then realize AVHE  in  hexagonal AFM monolayer.
The  proposed  way can produce large spin splitting around the
Fermi energy level, which is due to $d$ orbital mismatch among these magnetic atoms.
  By first-principles calculations, we translate our proposed way  into $\mathrm{CrMoC_2S_6}$ monolayer and clarify the isovalent alloying assisted  AVHE in hexagonal AFM monolayer.

\section{Ways to achieve AVHE}
The isovalent alloying assisted  AVHE in hexagonal AFM monolayer is schematically illustrated in \autoref{sy}.
\autoref{sy} (a) shows that the two magnetic atoms of a hexagonal AFM  monolayer   per unit cell are the same 3$d$ elements, and its  energy extrema of conduction/valance bands is at high symmetry -K/K point.  We assume that the lattice of hexagonal AFM  monolayer has spatial inversion symmetry, but the AFM ordering  leads to the simultaneously broken inversion ($P$) symmetry and time reversal ($T$)
symmetry, which produces spontaneous valley polarization (\autoref{sy} (c)). However, the system possesses $PT$ symmetry, which gives rise to spin degeneracy, prohibiting the realization of AVHE. Due to $PT$ symmetry, the zero berry curvature ($\Omega(k)$) everywhere in the momentum space can be observed. However, the Berry curvatures for the spin-up and
spin-down channels are nonzero, and they are equal in magnitude and opposite in sign.
In the presence of a longitudinal in-plane electric field,  an appropriate electron doping  produces  valley-spin Hall effect  (\autoref{sy} (e)).

Here, substituting elements  via isovalent alloying (One of the two magnetic atoms is replaced by a 4$d$ element with the same outer electrons.) can be applied to break the $PT$ symmetry(\autoref{sy} (b)), and the spin
degeneracy of -K and K valleys can be removed.   Our proposed system can be called asymmetric AFM semiconductor (AAFMS)\cite{v16}.
Although the two magnetic atoms (3$d$  and 4$d$ elements)  have the same  surrounding atomic arrangement, the energy level of 4$d$ electron is higher than that of 3$d$ one ($d$ orbital mismatch), producing spin splitting.
Due to broken $PT$ symmetry, the  spin
splitting of the -K and K valleys can be observed (\autoref{sy} (d)), resulting in the AVHE  (\autoref{sy} (f)).

For bulk materials, the AAFMS has been predicted in  a family
of double perovskites $\mathrm{A_2CrMO_6}$ (A = Ca, Sr, Ba, and M = Ru, Os)\cite{v16}. The large spin polarization around the
Fermi energy level has been observed due to $d$ orbital mismatch among these magnetic ions.
For 2D materials, $\mathrm{CrMoA_2S_6}$ (A = C, Si, or Ge) monolayers are AAFMSs,   $\mathrm{CrMoC_2S_6}$ of which  has
the highest N$\acute{e}$el temperature  of 556 K with robust magnetism against carrier doping and external in-plane
strain\cite{v17}. Next, we take  $\mathrm{CrMoC_2S_6}$ monolayer as an example to illustrate our proposal.

\begin{figure}
  \includegraphics[width=8cm]{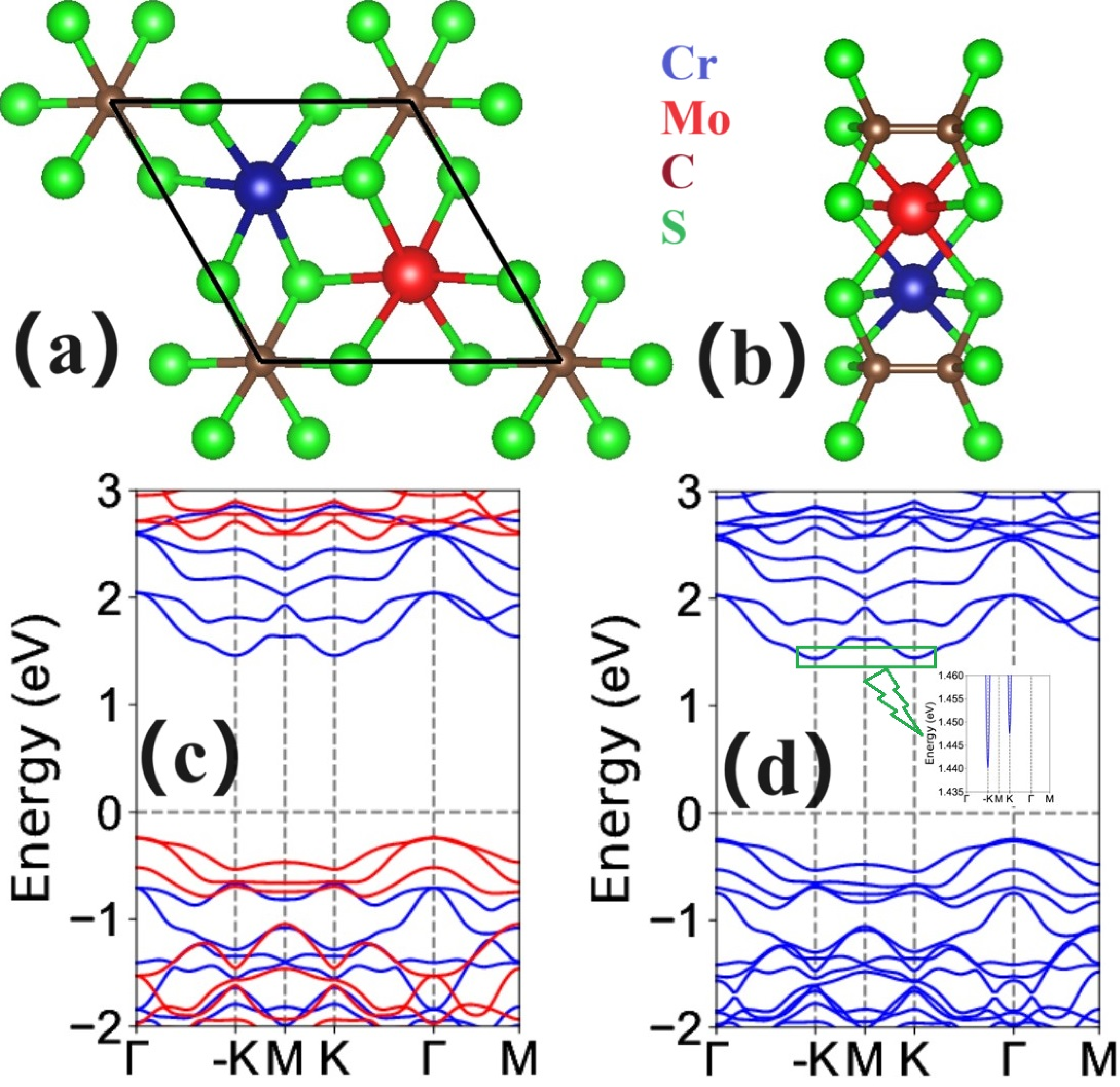}
  \caption{(Color online) For monolayer  $\mathrm{CrMoC_2S_6}$,  (a) and (b): the top and side views of crystal structures; (c) and (d): the energy band structures without SOC and with SOC. In (c), the spin-up and spin-down channels are depicted in blue and red. In (d), the inset shows the enlarged portion of conduction bands near the Fermi energy level.}\label{st}
\end{figure}

\section{Computational detail}
We perform the spin-polarized  first-principles calculations within density functional theory (DFT)\cite{1} by using the projector augmented-wave (PAW) method,  as implemented in Vienna ab initio Simulation Package (VASP)\cite{pv1,pv2,pv3}.  We use the generalized gradient
approximation  of Perdew-Burke-Ernzerhof (PBE-GGA)\cite{pbe}as the exchange-correlation functional.
The kinetic energy cutoff  of 500 eV,  total energy  convergence criterion of  $10^{-8}$ eV, and  force convergence criterion of 0.0001 $\mathrm{eV.{\AA}^{-1}}$  are set to obtain the accurate results.
To describe the strong correlation interaction of 3$d$ and 4$d$ electrons in  $\mathrm{CrMoC_2S_6}$, a Hubbard correction $U_{eff}$=3.0 eV\cite{v17}  is used by  the
rotationally invariant approach proposed by Dudarev et al\cite{u}. The SOC is incorporated for investigation of valley splitting and magnetic anisotropy energy (MAE).
To avoid the interaction between
adjacent layers, the vacuum space of more than 20 $\mathrm{{\AA}}$ along $z$ direction is adopted.
  A 12$\times$12$\times$1 $\Gamma$-centered Monkhorst-Pack grid is used to sample the Brillouin zone (BZ).
Based on  finite displacement method, the interatomic force constants (IFCs)  are calculated by employing 3$\times$3$\times$1 supercell, and the phonon dispersion spectrum  is obtained by the  Phonopy code\cite{pv5}. The elastic stiffness tensor  $C_{ij}$   are calculated by using strain-stress relationship (SSR) method.  The  $C^{2D}_{ij}$ has been renormalized by   $C^{2D}_{ij}$=$L_z$$C^{3D}_{ij}$, where the $L_z$ is  the length of unit cell along $z$ direction.
The Berry curvatures
are calculated directly from the calculated
wave functions  based on Fukui's
method\cite{bm},  as implemented in  the VASPBERRY code\cite{bm1,bm2,bm3}.
\begin{figure}
  \includegraphics[width=7.0cm]{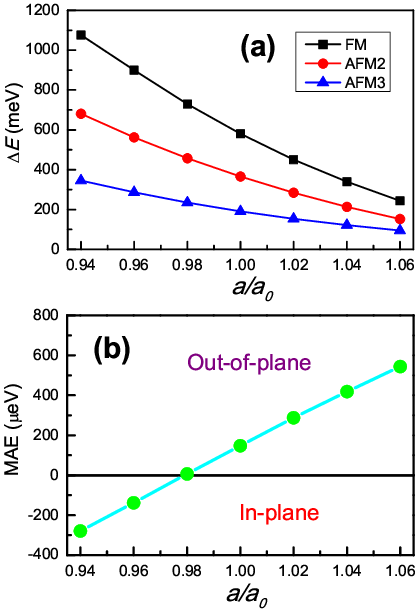}
\caption{(Color online)For  $\mathrm{CrMoC_2S_6}$, (a): the energy difference between  FM/AFM2/AFM3 and AFM1 orderings as a function of $a/a_0$; (b): the MAE vs $a/a_0$.}\label{ene}
\end{figure}

\begin{figure}
  \includegraphics[width=8cm]{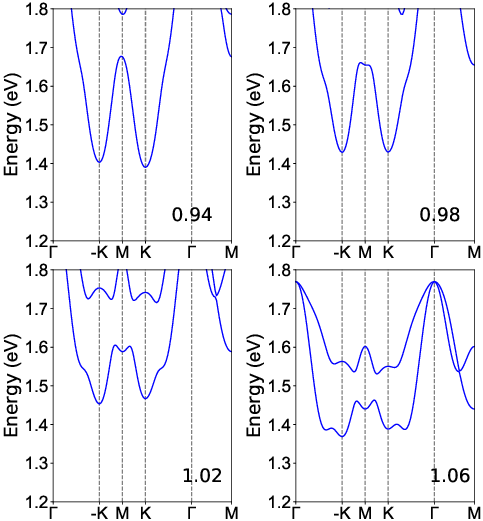}
\caption{(Color online)For  $\mathrm{CrMoC_2S_6}$, the energy band structures of conduction bands near the Fermi energy level with SOC at representative $a/a_0$ (0.94, 0.98, 1.02, 1.06).}\label{band}
\end{figure}

\section{Crystal and electronic structures}
It has been proved that  $\mathrm{CrMoC_2S_6}$ monolayer possesses dynamical and thermal stabilities, and the experimental possibility is also confirmed by the formation energy\cite{v17}. Here, we also calculate two independent elastic constants: $C_{11}$=91.69 $\mathrm{Nm^{-1}}$ and $C_{12}$=34.57  $\mathrm{Nm^{-1}}$,  which  satisfy the  Born  criteria of mechanical stability\cite{ela}:
$C_{11}>0$ and $C_{11}-C_{12}>0$,  confirming  its mechanical stability.
The crystal structures of  $\mathrm{CrMoC_2S_6}$  are  shown in \autoref{st} (a) and (b), which crystallizes in the  $P312$ space group (No.149),  lacking spatial inversion symmetry.  The $\mathrm{CrMoC_2S_6}$ can be obtained by substituting one Cr of  $\mathrm{Cr_2C_2S_6}$ with Mo via isovalent alloying. The Cr and Mo are surrounded by six S
atoms, forming a honeycomb lattice. Two $\mathrm{CS_3}$ moieties are connected by two C atoms, forming a dumbbell-like
structure. The optimized  equilibrium lattice constants of $\mathrm{CrMoC_2S_6}$  are $a$=$b$=5.714 $\mathrm{{\AA}}$ by GGA+$U$ method.
To determine magnetic ground state of  $\mathrm{CrMoC_2S_6}$,  the FM and three  AFM configurations (AFM1, AFM2 and AFM3) are constructed  (FIG.S1 of electronic supplementary information (ESI)), and the AFM1  is called N$\acute{e}$el AFM state.
  The energy of AFM1 per unit cell is 580 meV, 365 meV and  191 meV  lower  than those of FM, AFM2 and AFM3 cases, confirming that the $\mathrm{CrMoC_2S_6}$ possesses AFM1 ground state.

 The energy band structures of  $\mathrm{CrMoC_2S_6}$  and $\mathrm{Cr_2C_2S_6}$ without SOC are plotted in \autoref{st} (c) and FIG.S2 of ESI, respectively. The lattice of $\mathrm{Cr_2C_2S_6}$ has  spatial inversion symmetry, but the  inversion symmetry will disappear with AFM1 ordering. However,  a combination of inversion symmetry $P$ and time-reversal symmetry $T$  ($PT$) exist, which leads to disappeared spin splitting.
 For $\mathrm{CrMoC_2S_6}$, the large spin splitting around the
Fermi energy level can be observed, and the  total magnetic moment per unit cell is strictly 0.00 $\mu_B$. It is clearly seen that  $\mathrm{CrMoC_2S_6}$ is an indirect gap semiconductor with  conduction band bottom (CBM)/valence band maximum (VBM) at K/$\Gamma$ point, and the energies of  -K and K valleys in the conduction bands are degenerate. Monolayer $\mathrm{CrMoC_2S_6}$ is  a bipolar AFM semiconductor with VBM and
CBM in different spin channels.

When including SOC, the magnetic orientation of $\mathrm{CrMoC_2S_6}$  should be  determined by MAE, and only out-of-plane case can produce spontaneous valley polarization\cite{gsd1}. The MAE  can be calculated  by $E_{MAE}=E^{||}_{SOC}-E^{\perp}_{SOC}$, where $||$ and $\perp$  mean that spins lie in the plane and out-of-plane. The positive MAE means out-of-plane magnetic anisotropy, while the negative value indicates in-plane one. The predicted MAE is 148$\mathrm{\mu eV}$/unit cell, which means the out-of-plane easy magnetization axis of $\mathrm{CrMoC_2S_6}$. The energy band structures of  $\mathrm{CrMoC_2S_6}$  with SOC are plotted in \autoref{st} (d), which  shows  the  valley polarization with the valley splitting of  only 7.4  meV ($\Delta E_C=E_{K}^C-E_{-K}^C$), and  the energy of K valley
is higher than one of -K valley.
In order to strengthen valley splitting, strain engineering may be an effective and feasible method\cite{gd}.
In the next section, we will study the strain effects on magnetic and electronic properties of $\mathrm{CrMoC_2S_6}$.

\begin{figure}
  \includegraphics[width=8cm]{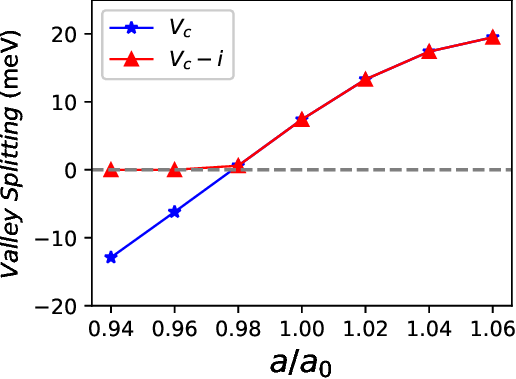}
\caption{(Color online)For $\mathrm{CrMoC_2S_6}$, the valley splitting with  assumed out-of-plane  and intrinsic magnetization direction  ($V_c$ and $V_c-i$) for conduction  band   as a function of $a/a_0$. }\label{ss}
\end{figure}

\section{strain effects}
Here, the $a/a_0$ (0.94$\sim$1.06) is used to simulate the biaxial strain, where   $a$ and $a_0$ are the strained and unstrained lattice constants, respectively. The $a/a_0$$<$1/$>$1 means the compressive/tensile strain. Firstly, we determine the magnetic ground state of strained  $\mathrm{CrMoC_2S_6}$ by the
energy difference between  FM/AFM2/AFM3 and AFM1 configurations, which is shown in \autoref{ene} (a).
Within considered $a/a_0$ range,    the AFM1 ordering is always ground state. With increasing strain, the strength of the magnetic interaction is reduced. To maintain spontaneous valley polarization, the direction of magnetization is another key factor.
 The MAE vs $a/a_0$ is plotted in \autoref{ene} (b), which shows a transition of  magnetization direction.
 Compressive strain can make the magnetization direction of $\mathrm{CrMoC_2S_6}$ change from out-of-plane case to in-plane case, and this critical strain point is about 0.98. Fortunately,  tensile strain can  enhance the MAE of $\mathrm{CrMoC_2S_6}$, thereby promoting the spontaneous valley polarization.

\begin{figure}
  \includegraphics[width=8cm]{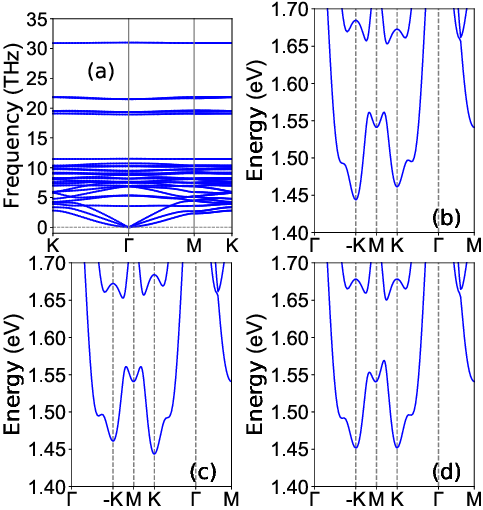}
\caption{(Color online)For  $\mathrm{CrMoC_2S_6}$ with $a/a_0$=1.04, the phonon dispersion curves (a); the energy band structures of conduction bands near the Fermi energy  level with SOC (b, c, d) for magnetization direction along the positive $z$, negative $z$, and positive $x$ direction, respectively.}\label{band-1}
\end{figure}
\begin{figure*}
  \includegraphics[width=12cm]{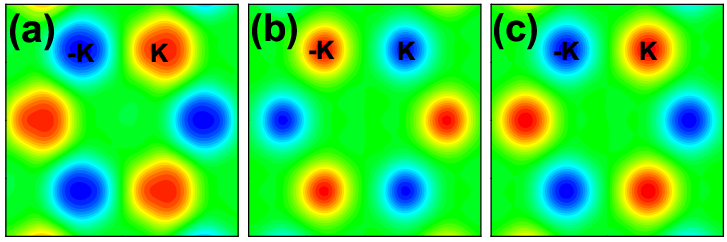}
\caption{(Color online)For  $\mathrm{CrMoC_2S_6}$ with $a/a_0$=1.04, the distribution of Berry curvatures of  total (a),  spin-up (b) and
spin-down (c).}\label{berry}
\end{figure*}

The energy band structures of $\mathrm{CrMoC_2S_6}$ at representative $a/a_0$ by using both GGA and GGA+SOC are plotted in FIG.S3 of ESI, and the  enlarged figures of SOC energy band structures near the Fermi energy  level for the conduction band are plotted in \autoref{band}.
The valley splitting with  assumed out-of-plane  and intrinsic magnetization direction   for conduction band   as a function of $a/a_0$ are plotted in \autoref{ss}. Based on FIG.S3 of ESI,  $\mathrm{CrMoC_2S_6}$ is always a bipolar AFM semiconductor. Near the Fermi energy level, the
the conduction bands comes from
spin-up channel, while the valence band
is dominated by spin-down channel.  With assumed out-of-plane magnetization, the compressive strain can  induce K valley polarization, while  the tensile strain maintains -K valley polarization. Both increasing compressive and tensile strains can enhance the absolute size of valley splitting $|\Delta E_C|$. From this perspective, strain can be considered analogous to a pseudomagnetic field.
 Stain-driven valley polarization transition can also be observed in Janus  $\mathrm{GdClF}$ monolayer\cite{gsd3}.
When considering intrinsic magnetic anisotropy, the spontaneous valley polarization appears with $a/a_0$ being larger than 0.98.
At $a/a_0$=1.06, the corresponding valley splitting is about 19.5  meV.

Finally, taking a typical $a/a_0$=1.04, the physical properties of  $\mathrm{CrMoC_2S_6}$ are detailedly investigated.
The calculated phonon spectrum  of   $\mathrm{CrMoC_2S_6}$  with $a/a_0$=1.04   is shown  in \autoref{band-1} (a), which shows  no obvious imaginary frequencies,  indicating its dynamic stability.
The energy band structures of conduction bands near the Fermi energy level with SOC for magnetization direction along the positive $z$, negative $z$, and positive $x$ directions are plotted in \autoref{band-1} (b), (c) and (d).
\autoref{band-1} (b) shows the  valley polarization with the valley splitting of 17.4 meV, and  the energy of K valley
is higher than one of -K valley. \autoref{band-1} (c)  indicates that the valley polarization can  be
switched by reversing the magnetization direction. When the  magnetization direction of $\mathrm{CrMoC_2S_6}$ is in-plane  (see \autoref{band-1} (d)), no valley polarization can be observed.

\begin{table}
\centering \caption{ For  monolayer  $\mathrm{CrMoC_2S_6}$ with $a/a_0=1.04$ at different $U$ values,   the energies of FM, AFM1, AFM2 and AFM3 ordeing (meV);  the MAE/unit cell in $\mathrm{\mu eV}$; the valley splitting ($V_c$) for conduction band (meV). }\label{tab}
  \begin{tabular*}{0.48\textwidth}{@{\extracolsep{\fill}}ccccccc}
  \hline\hline
$U$ (Cr, Mo)& $E_{FM}$ &  $E_{AFM1}$& $E_{AFM2}$& $E_{AFM3}$& MAE&$V_{c}$\\\hline\hline
 (3,2)&371&0&231&128&478&16.5\\\hline
 (3,3)&339&0&214&122&419&17.4 \\\hline
 (4,3)&292&0&181&107&416&18.2\\\hline\hline
\end{tabular*}
\end{table}

For $\mathrm{CrMoC_2S_6}$ at $a/a_0$=1.04, the distribution of Berry curvatures of  total,  spin-up  and
spin-down  are plotted in \autoref{berry}. For purposes of comparison, the related distribution of Berry curvatures of  $\mathrm{Cr_2C_2S_6}$
are also  shown in FIG.S4 of ESI. Due to $PT$ symmetry, the  total Berry curvatures of  $\mathrm{Cr_2C_2S_6}$ are zero, but the  berry curvatures are opposite  for the same valley at different spin channel and different valley at the same spin channel.
However, for $\mathrm{CrMoC_2S_6}$, the total Berry curvatures are nonzero at -K and K valley due to broken $PT$ symmetry.  The Berry curvatures
 of the -K and K valleys have opposite signs for the same spin channel,  and the Berry curvatures are also opposite  for the same valley at different spin channel.
 With an applied longitudinal in-plane electric field,
the Bloch carriers will acquire an anomalous transverse
velocity $v_{\bot}$$\sim$$E_{\parallel}\times\Omega(k)$\cite{q4}. When
 the Fermi energy level is shifted between the -K and K valleys in the conduction band, the spin-up carriers from -K valley will
accumulate along one edge of the sample, resulting
in the AVHE (\autoref{sy} (f)).

In general, 3$d$ electrons have a stronger electron correlation than 4$d$ electrons.
For  monolayer  $\mathrm{CrMoC_2S_6}$ at $a/a_0=1.04$, the different $U$ (Cr, Mo) values are also considered to confirm the reliability of our results. From $U$ (3, 2) to $U$ (3, 3) to $U$ (4, 3), the magnetic interaction energy and MAE of $\mathrm{CrMoC_2S_6}$ become weak, but it always possesses out-of-plane AFM1 ordering.  At different $U$ (Cr, Mo) values, the energy band structures of conduction bands near the Fermi energy level
with SOC for magnetization direction along the positive $z$ direction are plotted in FIG.S5 of ESI. They all show -K valley polarization, and the valley splitting increases with increasing  $U$ (Cr, Mo). The related data are summarized in \autoref{tab}.

To elucidate
mechanical performance of $\mathrm{CrMoC_2S_6}$, the isotropic in-plane Young's modulus $C_{2D}$ of 78.66 $\mathrm{Nm^{-1}}$ is predicted. This   is  weaker than graphene ($\sim 340\pm 40$ Nm$^{-1}$) and MoS$_2$ ($\sim 126.2$ Nm$^{-1}$)~\cite{q5-1,q5-1-1}, indicating the better mechanical flexibility of monolayer   $\mathrm{CrMoC_2S_6}$, which  is in favour of experimentally enhancing valley splitting by strain.  For $\mathrm{MoS_2}$ monolayer, the  strain with  as large as 5.6\% has been achieved experimentally\cite{q5-1-2}. Therefore, the   1.04 (4\%) strain in $\mathrm{CrMoC_2S_6}$ can be realized  experimentally due to small  $C_{2D}$.

\section{Conclusion}
In summary,  we present a way to induce AVHE in hexagonal AFM monolayer by isovalent alloying.
By the first-principles calculations, we confirm the validity of our proposed way  by an extensive study of $\mathrm{CrMoC_2S_6}$ monolayer.
Due to broken $P$ and $T$ symmetries in $\mathrm{CrMoC_2S_6}$, there is a spontaneous valley polarization.
Moreover, the spin splitting can occur due to broken $PT$ symmetry, which can also be explained by $d$ orbital mismatch of different magnetic atoms. Strain engineering can be used to tune valley splitting, and calculated results show that  tensile strain  can enhance the valley splitting of $\mathrm{CrMoC_2S_6}$ with the out-of-plane AFM1 ordering.  With an applied longitudinal in-plane electric field,
the Berry curvature can produce AVHE. Our works
are expected to speed up the realization of AVHE  in AFM monolayers, and provide a
route for constructing energy-efficient and ultrafast valleytronic devices.

\begin{acknowledgments}
This work is supported by Natural Science Basis Research Plan in Shaanxi Province of China  (2021JM-456). We are grateful to Shanxi Supercomputing Center of China, and the calculations were performed on TianHe-2.
\end{acknowledgments}

\end{document}